\documentclass[a4page, 10pt]{article}
\usepackage{amssymb,amsmath}
\usepackage{graphicx}
\numberwithin{equation}{section}

\begin{document}

\title{Black Hole in a Heat Bath}

\author{Jarmo M\"akel\"a\footnote{Vaasa University of Applied Sciences,
Wolffintie 30, 65200 Vaasa, Finland, email: jarmo.makela@vamk.fi}}

\maketitle

\begin{abstract}

It was recently pointed out by Barrau, Martineau and Renevey that the mass of the Schwarzschild black hole immersed in a heat bath tends to infinity in a finite time. We show that this singularity problem in the black hole mass may be avoided, if the stretched horizon of the black hole is constructed out of a fixed, finite number of discrete constituents, and an appropriate counting is applied to the energy eigenstates of the hole.
 
\end{abstract}

\section{Introduction}

The thermal properties of black holes possess a subtle  and an intriguing feature, which is often overlooked, but was recently brought to a central stage by Barrau, Martineau and Renevey in Ref. \cite{yy}. As it is well known, the energy density of thermal radiation is proportional to the fourth power of its absolute temperature $T$. The amount of heat absorbed by a black hole in a unit of time, in turn, is proportional to its effective area which, for the Schwarzchild black hole, is proportional to the square of its Schwarzschild mass $M$. Hence one expects the mass of the Schwarzschild black hole immersed in a heat bath with temperature $T$ to evolve, from the point of view of a distant observer at rest with respect to the hole, according to an equation:
\begin{equation}
\frac{dM}{dt} = \gamma M^2T^4,
\end{equation}
where $\gamma$ is an appropriate positive constant. In Ref. \cite{yy} the effective radius of the Schwarzschild black hole was taken to be, in the natural units, $3\sqrt{3}M$, and the authors found that Eq. (1.1) takes the form:
\begin{equation}
\frac{dM}{dt} = \frac{36}{5}\pi^2 M^2T^4
\end{equation}
or, in the SI units:
\begin{equation}
\frac{dM}{dt} = \frac{36}{5}\pi^2\frac{k_B^4G^2}{\hbar^3c^8}M^2T^4.
\end{equation}

Eq. (1.3) may be easily integrated, and it implies:
\begin{equation}
\frac{1}{M_i} - \frac{1}{M} = \frac{36}{5}\pi^2\frac{k_B^4G^2}{\hbar^3c^8}t,
\end{equation}
where $M_i$ is the initial mass of the black hole, when the time $t=0$. One finds from Eq. (1.4) that the black hole mass $M$ tends to infinity in a {\it finite  time}
\begin{equation}
t_0 = \frac{5}{36\pi^2}\frac{\hbar^3c^8}{k_B^4G^2}\frac{1}{M_iT^4}
\end{equation}
Although this result, based on an approximation of a real, growing black hole by means of a Schwarzschild black hole, has little astrophysical relevance - even for a supermassive black hole with mass $M_i$ equal to around $10^9$ solar masses, or $10^{39}kg$, one finds that for $3K$ background radiation $t_0\sim 10^{35}s$ - an existence of singularity in the mass of a black hole immersed in a heat bath certainly presents an important problem. There are some grounds to believe that at the Planck scale one encounters with huge, Planck-scale temperatures. At such temperatures a Planck-size black hole increases infinitely large in a Planck time. Any singularity in a theory of physics signals its breakdown. Is there any way to get around of this problem?

   The problem of the singularity in the black hole mass received a detailed investigation in Ref. \cite{yy} (see also the references there). However, the treatment in Ref. \cite{yy} was mostly classical. In this paper we shall attempt a quantum-mechanical approach. Unless otherwise stated, we shall always use the natural units, where $\hbar=c=G=k_B = 1$.

\section{The Model}

 As in Refs. \cite{kaa,koo,nee,vii,kuu} we consider the statistical and the thermodynamical properties of the Schwarzschild black hole from the point of view of an observer with constant proper acceleration $a$, just outside of the event horizon of the hole. For an observer with constant radial coordinate $r$ in the Schwarzschild spacetime equipped with the Schwarzschild line element
\begin{equation}
ds^2= -\left(1 - \frac{2M}{r}\right)\,dt^2 + \frac{dr^2}{1 - \frac{2M}{r}} + r^2\,d\theta^2 + r^2\sin^2(\theta)\,d\phi^2
\end{equation}
the modulus of the proper acceleration vector field $a^\mu$ is \cite{kaa}
\begin{equation}
a = \left(1 - \frac{2M}{r}\right)^{-1/2}\frac{M}{r^2}.
\end{equation}
We shall assume that when the mass $M$ of the Schwarzschild black hole is changed, the radial coordinate $r$ of the observer is also changed, but in such a way that the proper acceleration $a$ of the observer is kept unchanged. In other words, we shall assume that when the mass $M$ takes on a small change $dM$, the radial coordinate $r$ of the observer takes on a change $dr$ such that
\begin{equation}
da = \frac{\partial a}{\partial M}\,dM + \frac{\partial a}{\partial r}\,dr = 0.
\end{equation}
For this kind of an observer the changes in the observed properties of the black hole reflect changes in its true, physical properties, rather than the changes in the frame of reference of the observer.

   As in Refs. \cite{kaa, koo, nee, vii} our idea is to construct a stretched horizon of the Schwarz\-schild black hole, where the proper acceleration $a$ of an oberver at rest with respect to the stretched horizon is kept as a constant, out of a {\it finite number of discrete constituents}. The number of these constituents, which is assumed to be very large, we shall denote by $N$. In the simplest possible model of this kind we write the area of the stretched horizon as:
\begin{equation}
A = \alpha(n_1 + n_2 + \cdots + n_N),
\end{equation}
where the quantum numbers $n_1, n_2,\dots, n_N$ are non-negative integers, and $\alpha$ is a positive constant to be determined later. In other words, we shall assume that each constituent $j=1, 2,\dots, N$ contributes to the stretched horizon an area, which is of the form $n_j\alpha$. The constituent is in {\it vacuum}, if $n_j=0$; otherwise it is in an {\it excited state}.

   An advantage of our model is that with an appropriate counting of states the partition function
\begin{equation}
Z(\beta) = \sum_n e^{-\beta E_n}
\end{equation}
of the black hole, where we have summed over the energy eigenstates $n$ of the hole, may be calculated explicitly. It has become firmly established (see Refs. \cite{kaa, koo, nee, vii, kuu, seite, kasi}) that from the point of view of an observer at rest on the stretched horizon with constant proper acceleration $a$ the quantity
\begin{equation}
E = \frac{a}{8\pi}A
\end{equation}
is a valid concept of energy for a black hole. As in Refs. \cite{kaa, koo, nee, vii} we sum over the excited states of the constituents of the stretched horizon. Moreover, we shall assume that whenever the quantum states of the constituents are permuted, the overall quantum state of the black hole will also change. With this counting of states the partition function of the hole takes, from the point of view of our observer, the form: \footnote{The counting of states is similar to the one used in Ref. \cite{kuu}}
\begin{equation}
\begin{split}
Z(\beta) =&\sum_{n_1=1}^\infty\exp\left(-\beta\alpha n_1\frac{a}{8\pi}\right)\\
               &+\sum_{n_1=1}^\infty\sum_{n_2=1}^\infty\exp\left(-\beta\alpha n_1\frac{a}{8\pi}\right)\exp\left(-\beta\alpha n_2\frac{a}{8\pi}\right)\\
               &+\cdots\\
               &+\sum_{n_1=1}^\infty\cdots\sum_{n_N=1}^\infty\exp\left(-\beta\alpha n_1\frac{a}{8\pi}\right)\cdots\exp\left(-\beta\alpha n_N\frac{a}{8\pi}\right).
\end{split}
\end{equation}
In the first term on the right hand side of Eq. (2.7) we have assumed that just one of the constituents is in an excited state, and we have summed over those states. In the second term we have assumed that two constituents are in excited states. Finally, in the last term we have assumed that all $N$ constituents are in excited states. 

   Defining the characteristic temperature
\begin{equation}
T_C := \frac{\alpha a}{8\pi\ln(2)}
\end{equation}
we may write the partition function as:
\begin{equation}
Z(\beta) = y(\beta) + [y(\beta)]^2 +\cdots + [y(\beta)]^N,
\end{equation}
where we have denoted:
\begin{equation}
y(\beta) := \sum_{n=1}^\infty 2^{-n\beta T_C} = \frac{1}{2^{\beta T_C} - 1}.
\end{equation}
On the right hand side of Eq. (2.9) we have just a geometrical sum with $N$ terms. We have:
\begin{equation}
Z(\beta) = \frac{y(\beta)}{1 - y(\beta)}\left\lbrace 1 - [y(\beta)]^N\right\rbrace,
\end{equation}
which gives for the partition function the final expression:
\begin{equation}
Z(\beta) = \frac{1}{2^{\beta T_C} - 2}\left[1 - \left(\frac{1}{2^{\beta T_C} - 1}\right)^N\right].
\end{equation}
The result is identical to the one obtained in Ref. \cite{kaa}.\footnote{In Eq. (3.14) of Ref. \cite{kaa} a minor calculational error brought $N+1$, instead of $N$, to the exponential. However, since $N$ is assumed to be very large, this does not change the implications of the calculations.} Eq. (2.12) holds, whenever the temperature $T=\frac{1}{\beta}\ne T_C$. If $T=T_C$ we have:
\begin{equation}
Z(\beta) = N.
\end{equation}

\section{Phase Transition and the Hawking Effect}

  Employing Eqs. (2.12) and (2.13) one finds that from the point of view of our observer the average energy per constituent is:
\begin{equation}
\begin{split}
\bar{E}(\beta) &:= -\frac{1}{N}\frac{\partial}{\partial\beta}\ln(Z(\beta))\\
                     &= \left[\frac{1}{N}\frac{2^{\beta T_C}}{2^{\beta T_C} - 2} - \frac{2^{\beta T_C}}{(2^{\beta T_C} - 1)^{N+1} - 2^{\beta T_C} + 1}\right]T_C\ln(2).
\end{split}
\end{equation}
As it was found in Ref. \cite{kaa}, 
\begin{equation}
\lim_{N\rightarrow 0}\bar{E}(\beta) = 0,
\end{equation}
when $T< T_C$, whereas
\begin{equation}
\lim_{N\rightarrow\infty}\bar{E}(\beta) = \frac{2^{\beta T_C}}{2^{\beta T_C} - 1}T_C\ln(2),
\end{equation}
when $T>T_C$. One also finds that
\begin{equation}
\frac{d\bar{E}(\beta)}{dT}\bigg\vert_{T=T_C} = \frac{1}{6}(\ln(2))^2 N+ \mathcal{O}(1),
\end{equation}
where $\mathcal{O}(1)$ denotes the terms of the order $N^0$ or less. Hence one observes that the black hole performs, at the characteristic temperature $T_C$ defined in Eq. (2.8), a {\it phase transition}: When $T<T_C$, $\bar{E}(\beta)=0$, which means that the constituents of the stretched horizon are effectively in the vacuum, where $n_j=0$ for all $j=1, 2, \dots, N$, whereas at the characteristic temperature $T_C$ $\bar{E}(\beta)$ suddenly jumps. Putting $\beta=\frac{1}{T_C}$ in Eq. (3.3) we may deduce that the latent heat per constituent associated with this phase transition is
\begin{equation}
\bar{L} = 2T_C\ln(2).
\end{equation}
Defining the average of the quantum numbers $n_j$ as:
\begin{equation}
\bar{n}(\beta) := \frac{n_1 + n_2 + \cdots + n_N}{N} = \frac{\bar{E}(\beta)}{T_C\ln(2)},
\end{equation}
where the last equality follows from Eqs. (2.4), (2.6) and (2.8), we may infer that 
\begin{equation}
\bar{n}(\beta) = 2
\end{equation}
after the phase transition has been completed. In other words, the constituents of the stretched horizon jump, in effect, from the vacuum to the second excited states during the phase transition. 

      Since the constituents of the stretched horizon are effectively in the vacuum, when $T<T_C$, we may regard the characteristic temperature $T_C$ as the lowest possible temperature of the Schwarzschild black hole. Employing Eqs. (2.2) and (2.8) we observe that the lowest possible temperature of the hole is, from the point of view our observer just outside of the event horizon of the hole, where $r=2M$:
\begin{equation}
T_C = \left(1 - \frac{2M}{r}\right)^{-1/2}\frac{\alpha}{32\pi M\ln(2)}.
\end{equation}
The Tolman relation \cite{ysi} implies that from the point of view of a distant observer at rest with respect to the hole this corresponds to the temperature
\begin{equation}
T_{BH} = \frac{\alpha}{32\pi M\ln(2)}.
\end{equation}
If we choose
\begin{equation}
\alpha = 4\ln(2),
\end{equation}
we get:
\begin{equation}
T_{BH} = \frac{1}{8\pi M},
\end{equation}
which agrees with the {\it Hawking temperature} \cite{kymmenen}
\begin{equation}
T_H := \frac{1}{8\pi M}
\end{equation}
of the Schwarzschild black hole. Hence our simple model predicts, with the choice (3.10) for the parameter $\alpha$, the Hawking effect: Black hole is not entirely black, but it emits thermal radiation with a characteristic temperature, which equals with its Hawking temperature.

\section{Mass vs. Temperature}

A really interesting result is Eq. (3.3), which gives the average energy of a constituent, when the black hole is in a heat bath with temperature higher than its Hawking temperature. Eq. (3.3) implies, in the limit, where the number $N$ of the constituents tends to infinity:
\begin{equation}
T = -\frac{T_C\ln(2)}{\ln\left[1 - \frac{T_C\ln(2)}{\bar{E}}\right]}
   = -\frac{T_C\ln(2)}{\ln\left(1 - \frac{1}{\bar{n}}\right)}
   = -\frac{T_C\ln(2)}{\ln\left[1 - \frac{4N\ln(2)}{A}\right]}
\end{equation}
where we have used Eq. (3.6) in the second equality, and Eqs. (2.4), (3.6) and (3.10) in the third equality. Since our observer is just outside of the event horizon, we may identify, for all practical purposes, the stretched horizon area $A$ with the event horizon area $4\pi(2M)^2 = 16\pi M^2$ of the Schwarzschild black hole. Moreover, if we consider the thermodynamics of the black hole from the point of view of a faraway observer, instead of an observer just outside of the event horizon, we must replace the characteristic temperature $T_C$ by the Hawking temperature $T_H$ of the hole. This gives us the following relationship between the Schwarzschild mass $M$ of the hole, and the temperature $T_{BH}$ measured at the faraway infinity:
\begin{equation}
T_{BH} = -\frac{1}{8\pi M}\frac{\ln(2)}{\ln\left[1 - \frac{N\ln(2)}{4\pi M^2}\right]}.
\end{equation}

   When writing Eq. (4.2) we have assumed that the black hole has already performed the phase transition, and therefore, according to Eq. (3.7), $\bar{n} \ge 2$. In the highly excited states
\begin{equation}
\bar{n} = \frac{4\pi M^2}{N\ln(2)} \gg 1,
\end{equation}
and using the result:
\begin{equation}
\ln(1+x) = x - \frac{1}{2}x^2 + \frac{1}{3}x^3 - \cdots
\end{equation}
we get:
\begin{equation}
T_{BH} = \frac{M}{2N} + \mathcal{O}(\frac{1}{M}),
\end{equation}
Where $\mathcal{O}(\frac{1}{M})$ denotes the terms proportional to the first or higher powers of $\frac{1}{M}$. So we find that when the constituents of the Schwarzschild black hole are, in average, in highly excited states, we may write, in effect:
\begin{equation}
M = 2NT_{BH}
\end{equation}
or, in the SI units:
\begin{equation}
M = 2\frac{k_B}{c^2}NT_{BH}.
\end{equation}

\section{Black Hole in a Heat Bath}

As one may observe from Eq. (4.7), at high temperatures the Schwarzschild mass $M$ of the Schwarzschild black hole {\it grows linearly as a function of its temperature $T_{BH}$}. This effect was brought along by our assumption that the stretched horizon of the Schwarzschild black hole has a fixed, finite number $N$ of discrete constituents. So we find that even though the Hawking temperature $T_H$ is the lowest possible temperature of the black hole, it is not, according to our model, its only possible temperature. When the hole is in a heat bath, whose temperature is higher than its Hawking temperature, the hole begins to absorb heat and, at the same time, its temperature increases. When the temperature is high enough, the mass $M$ of the hole increases, in effect, linearly as a function of its temperature $T_{BH}$.

The result obtained above provides a possible solution to the problem of the singularity in the mass of a black hole immersed in a heat bath. Since the black hole has a certain temperature $T_{BH}$, which depends on its mass $M$ and the number $N$ of its constituents, we should modify Eq. (1.2) to the form:
\begin{equation}
\frac{dM}{dt} = \frac{36}{5}\pi^2M^2(T_{HB}^4 - T_{BH}^4).
\end{equation}
In this equation $T_{HB}$ is the temperature of the heat bath, whereas $T_{BH}$ is the temperature of the black hole. As one may observe, $\frac{dM}{dt}=0$, when 
\begin{equation}
T_{BH}=T_{HB},
\end{equation}
which means that the hole has reached a thermal equilibrium with the heat bath. The black hole no more absorbs heat, and the increase in its mass stops. In other words, in the given temperature $T_{HB}$ of the heat bath there is a certain maximum $M_{max}$ for the black hole mass $M$, which depends on the number $N$ of the constituents of its horizon. According to Eq. (4.2) this maximum $M_{max}$ is the solution of the equation:
\begin{equation}
T_{HB} = - \frac{1}{8\pi M_{max}}\frac{\ln(2)}{\ln\left[1 - \frac{N\ln(2)}{4\pi M_{max}^2}\right]}.
\end{equation}
When the temperature $T_{HB}$ is very much higher than the Hawking temperature $T_H$ of the hole, we may write, in effect:
\begin{equation}
M_{max} = 2NT_{HB}
\end{equation}
or, in the SI units:
\begin{equation}
M_{max} = 2\frac{k_B}{c^2}NT_{HB}.
\end{equation}
So, there is no singularity in the black hole mass.

How long does it take, until the black hole has reached the thermal equilibrium with the heat bath? Using Eq. (4.6) we may write Eq. (5.1) in the high-temperature limit as:
\begin{equation}
\frac{dM}{dt} = \frac{36}{5}\pi^2M^2\left(T_{HB}^4 - \frac{1}{16N^4}M^4\right).
\end{equation}
This equation may be easily integrated by means of the formula:
\begin{equation}
\int\frac{dx}{x^2 - a^4x^6} = -\frac{1}{x} + \frac{a}{2}\tanh^{-1}(ax) - \frac{a}{2}\tan^{-1}(ax) + C,
\end{equation}
and we find:
\begin{equation}
\begin{split}
t = \frac{5}{36\pi^2}\frac{1}{T_{HB}^4}\big\lbrack&- \frac{1}{M} + \frac{1}{M_i} + \frac{1}{4NT_{HB}}\tanh^{-1}\left(\frac{M}{2NT_{HB}}\right)\\ 
                                                                      &-\frac{1}{4NT_{HB}}\tanh^{-1}\left(\frac{M_i}{2NT_{HB}}\right)\\
                                                                      &-\frac{1}{4NT_{HB}}\tan^{-1}\left(\frac{M}{2NT_{HB}}\right)\\
                                                                      &+\frac{1}{4NT_{HB}}\tan^{-1}\left(\frac{M_i}{2NT_{HB}}\right)\big\rbrack,
\end{split}
\end{equation}
where $M_i$ is the initial mass of the hole, when $t=0$. Because the function $\tanh^{-1}(x)$ has the property:
\begin{equation}
\lim_{x\rightarrow 1^-}[\tanh^{ -1}(x)] = \infty,
\end{equation}
we have:
\begin{equation}
\lim_{M\rightarrow M_{max}^-}(t) = \infty.
\end{equation}
This means that the black hole will never quite reach the thermal equilibrium with the heat bath, even though it gets closer and closer to the equilibrium all the time.

\section{Concluding Remarks}

In this paper we have shown that the problem of the singularity in the mass of a black hole immersed in a heat bath may be avoided by means of an assumption that the stretched horizon of the hole which, for all practical purposes, may be identified with the event horizon of the hole, consists of a fixed, finite number of discrete constituents. Our model implies that the black hole has a certain minimum temperature which, by means of an appropriate choice of the parameter of the model, agrees with the Hawking temperature of the hole. However, if the temperature of the heat bath exceeds the Hawking temperature of the hole, the hole begins to absorb heat, and the constituents jump to higher excited states. As a consequence, the Schwarzschild mass of the hole increases and, at the same time, the black hole heats up. When the time passes, the hole tends to a thermal equilibrium with the heat bath. During this process the Schwarzschild mass of the black hole tends towards a certain maximum value, which is finite, and depends both on the temperature of the heat bath, and on the number of the constituents of the horizon. So, there is no singularity in the black hole mass.

    The results of our model provide support to the idea of a fundamentally discrete structure of space. In this paper we used the simplest possible model, where the black horizon area spectrum was evenly spaced. The use of such a model enabled us to carry out the calculations explicitly. It appears to the author that qualitatively similar results may be obtained by means of a model, where the area spectrum is similar to the one used in loop quantum gravity. \cite{kuu, yytoist}

\end{document}